\global\def\draftcontrol{0}

   \def\versionno{ sound in magnetic }

\catcode`\@=11

\expandafter\ifx\csname draftcontrol\endcsname\relax\global\def\draftcontrol{0}
\fi

{\count255=\time\divide\count255 by 60
\xdef\hourmin{\number\count255}
\multiply\count255 by-60\advance\count255 by\time
\xdef\hourmin{\hourmin:\ifnum\count255<10 0\fi\the\count255}}
\def\draftdate{\number\month/\number\day/\number\year\ \ \ \hourmin }

\newcommand\makepapertitle{\par
  \begingroup
    \renewcommand\thefootnote{\@fnsymbol\c@footnote}%
    \def\@makefnmark{\rlap{\@textsuperscript{\normalfont\@thefnmark}}}%
    \long\def\@makefntext##1{\parindent 1em\noindent
            \hb@xt@1.8em{%
                \hss\@textsuperscript{\normalfont\@thefnmark}}##1}%
     \newpage
     \global\@topnum\z@   
     \@makepapertitle
     \thispagestyle{empty}\@thanks
  \endgroup
  \setcounter{footnote}{0}%
  \global\let\thanks\relax
  \global\let\makepapertitle\relax
  \global\let\@makepapertitle\relax
  \global\let\@thanks\@empty
  \global\let\@author\@empty
  \global\let\@date\@empty
  \global\let\@title\@empty
  \global\let\title\relax
  \global\let\author\relax
  \global\let\date\relax
  \global\let\and\relax
  \def\version{\let\version\@version\@gobble}
}
\def\@makepapertitle{%
  \newpage
   \ifnum\draftcontrol=1 {}
   \version\versionno
   \vskip 3em%
   \else
   \hfill\hbox to 3cm {\parbox{4cm}{\@pubnum}\hss}%
   \vskip 3em%
   \fi
   \begin{center}%
   \let \footnote \thanks
     {\LARGE {\@title}}%
     \vskip 1.5em%
     {\normalsize
       \lineskip .5em%
       \begin{tabular}[t]{c}%
         \@author
       \end{tabular}\par}%
     \vskip 1.5em%
     {\@bstract}%
     \end{center}%
     \vskip 1.5em
     \@date%
   \par
}

\gdef\@pubnum{}
\def\pubnum#1{%
  \gdef\@pubnum{#1}}

\gdef\@bstract{}
\def\Abstract#1{%
  \gdef\@bstract{%
   \parbox{\textwidth-0pc}{%
   \centerline{\bf Abstract}\penalty1000%
\kern.2cm%
\noindent
\renewcommand\baselinestretch{1.0}%
{#1}}}
}

\def\ps@paper{\let\@mkboth\@gobbletwo%
     \ifnum\draftcontrol=1
    \def\@oddfoot{\hbox to \textwidth{\tiny \versionno \hfil\tiny\draftdate}%
    \hskip -\textwidth \hbox to \textwidth{\hfil\rm\thepage\hfil}}%
     \else\def\@oddfoot{\hbox to \textwidth{\hfil\rm\thepage\hfil}}
     \fi
     \let\@evenfoot\@oddfoot
}

\def\body{\clearpage
          \pagestyle{paper}
    }

\def\@version#1{\ifnum\draftcontrol=1
\typeout{}\typeout{#1}\typeout{}
\vskip3mm\centerline{\hbox{\fbox{\normalsize{\tt DRAFT -- #1 -- }
                   {\draftdate}}}}\vskip3mm
\fi}
\let\version\@version
\long\def\eqlabel#1{\ifnum\draftcontrol=1
                    \tag@false  
                    \tag*{(\theequation) \hbox to -0.2cm{\hspace{0cm}\small{#1}\hss}}
                    \refstepcounter{equation}
                    \edef\@currentlabel{\theequation}
                    \ltx@label{#1}          
                    \else
                    \label{#1}
                    \fi
                    }
\let\st@bibitem\@bibitem
\let\st@lbibitem\@lbibitem
\ifnum\draftcontrol=1
  \def\@bibitem#1{%
    \st@bibitem{#1}\a@@label{#1}\ignorespaces}
  \def\@lbibitem[#1]#2{%
    \st@lbibitem[#1]{#2}\a@@label{#2}\ignorespaces}
  \def\a@@label#1{%
    \gdef\a@lab{\smash{\normalfont\small#1}}
    \ifvmode
      \if@inlabel
        \global\setbox\@labels\hbox{%
          \llap{\a@lab\let\a@lab\relax
                \kern\@totalleftmargin\kern\marginparsep}%
          \box\@labels}%
      \fi
    \fi}
\fi

\documentclass[12pt,letterpaper]{article}

\usepackage{amsmath,amssymb,array,calc,epsfig}

\ifnum\draftcontrol=1
\fi

\renewcommand\baselinestretch{1.25}
\renewcommand\section{\@startsection {section}{1}{\z@}%
                                   {-3.5ex \@plus -1ex \@minus -.2ex}%
                                   {2.3ex \@plus.2ex}%
                                   {\normalfont\large\bfseries}}
\renewcommand\subsection{\@startsection{subsection}{2}{\z@}%
                                   {-3.25ex\@plus -1ex \@minus -.2ex}%
                                   {1.5ex \@plus .2ex}%
                                   {\normalfont\normalsize\bfseries}}
\renewcommand\subsubsection{\@startsection{subsubsection}{3}{\z@}%
                                   {-3.25ex\@plus -1ex \@minus -.2ex}%
                                   {1.5ex \@plus .2ex}%
                                   {\normalfont\normalsize\it}}
\renewcommand\paragraph{\@startsection{paragraph}{4}{\z@}%
                                   {-3.25ex\@plus -1ex \@minus -.2ex}%
                                   {1.5ex \@plus .2ex}%
                                   {\normalfont\normalsize\bf}}

\numberwithin{equation}{section}

\def\ie{{\it i.e.}}

\def\revise#1       {\raisebox{-0em}{\rule{3pt}{1em}}%
                     \marginpar{\raisebox{.5em}{\vrule width3pt\
                     \vrule width0pt height 0pt depth0.5em
                     \hbox to 0cm{\hspace{0cm}{%
                     \parbox[t]{4em}{\raggedright\footnotesize{#1}}}\hss}}}}

\def\calc         {{\cal C}}

\def\cali         {{\cal I}}

\def\calo         {{\cal O}}

\def\zet          {{\mathbb Z}}

\def\del          {\partial}

\def\sqr#1#2{{\vcenter{\vbox{\hrule height.#2pt
 \hbox{\vrule width.#2pt height#1pt \kern#1pt
 \vrule width.#2pt}\hrule height.#2pt}}}}

\def\a{\alpha}
\def\b{\beta}
\newcommand{\qq}{\mathfrak{q}}
\newcommand{\ww}{\mathfrak{w}}

\def\e{\epsilon}
\def\r{\rho}

\def\w{\omega}
\def\hh{\hat{h}}
\def\ha{\hat{a}}

\newcommand{\be}{\begin{equation}}

\newcommand{\ee}{\end{equation}}
\newcommand{\bea}{\begin{eqnarray}}
\newcommand{\eea}{\end{eqnarray}}

\newcommand{\bra}{{\langle}}
\newcommand{\ket}{{\rangle}}

\catcode`\@=12

\begin{document}


\title{\bf Sound Waves in (2+1) Dimensional Holographic Magnetic Fluids}
\pubnum{UWO-TH-08/16
}

\date{October 2008}

\author{
Evgeny I. Buchbinder$ ^{1}$,  Alex Buchel$ ^{1,2}$ and Samuel E. V\'azquez$ ^{1}$\\[0.4cm]
\it $ ^1$Perimeter Institute for Theoretical Physics\\
\it Waterloo, Ontario N2J 2W9, Canada\\
\it $ ^2$Department of Applied Mathematics\\
\it University of Western Ontario\\
\it London, Ontario N6A 5B7, Canada
}

\Abstract{
We use the AdS/CFT correspondence to study propagation of sound waves in
strongly coupled (2+1) dimensional conformal magnetic fluids. Our
computation provides a nontrivial consistency check of the viscous
magneto-hydrodynamics of Hartnoll-Kovtun-M\"uller-Sachdev to leading order
in the external field. Depending on the behavior of the magnetic field in the
hydrodynamic limit, we show that it can lead to further attenuation of
sound waves in the (2+1) dimensional conformal plasma, or reduce the speed
of sound. We present both field theory and dual supergravity descriptions
of these phenomena. While to the leading order in momenta the dispersion of the 
sound waves obtained from the dual supergravity description 
agrees with the one predicted from field theory, we find a discrepancy at 
higher order. This suggests that further corrections to
HKMS magneto-hydrodynamics are necessary.
}

\makepapertitle

\body

\version\versionno


\section{Introduction}
Gauge theory/string theory correspondence of Maldacena~\cite{m9711, m2, ss}
proved to be a valuable tool in study dynamical 
processes in strongly coupled four dimensional gauge theory plasma. 
One of the most impressive contribution of string theory 
to the non-equilibrium plasma phenomena was the construction of relativistic 
conformal hydrodynamics \cite{Baier, shiraz},
with its emphasis on shortcomings of widely used second order hydrodynamics 
of M\"uller-Israel-Stewart (MIS) \cite{m,is}.    

It has been realized recently that string theory, and, in particular, dual holographic 
descriptions of strongly coupled 
(2+1) dimensional collective dynamics, might shed new light on longstanding problems in condensed matter
such as the theory of quantum phase transitions~\cite{Kovtun}, superfluidity~\cite{Basu, cm1}, 
and high-temperature 
superconductivity~\cite{cm11, cm2}. Our present paper is largely motivated by these ideas. 
Specifically, we would like 
to better understand conformal viscous hydrodynamics of strongly coupled (2+1) quantum field theories in external 
fields. Our starting point  is the first order dissipative 
magneto-hydrodynamics proposed by Hartnoll {\it et.al.} (HKMS) in \cite{Kovtun}. 
Much like four-dimensional 
MIS relativistic hydrodynamics, it is built on the idea of constructing an entropy current 
away from equilibrium plus
constitutive relations for dissipative currents.
The entropy current is constrained to have a positive divergence, an idea 
originally due to Landau and Lifshitz~\cite{LL}. We already know that this framework misses some important aspects of relativistic 
hydrodynamics at second order~\cite{Baier}, but it
is a well-motivated approximation. In this paper we would like to subject 
HKMS magneto-hydrodynamics\footnote{We postpone 
detailed analysis of conformal viscous hydrodynamics in external 
fields to our companion paper \cite{inprep}.} to consistency tests 
by extracting transport properties from the dispersion relation of the sound waves in 
strongly coupled $M2$-brane plasma and comparing 
them with the transport coefficients 
obtained from analysis of the current-current correlation functions \cite{HK,HH}.  

Our second motivation is to use the external field as a `dial' 
to control the (effective) speed of sound waves in strongly coupled 
plasma. Indeed, the speed of sound waves $c_s$ in conformal plasma and in 
the absence of external fields is determined by 
simple scale invariance, $c_s=1/\sqrt{2}$ in (2+1) conformal fluids, 
in units where the speed of light is unity. This result is universal
in relativistic (2+1) dimensional conformal hydrodynamics
without external fields. On the other hand, interesting (2+1) dimensional gapless systems, 
such as a single atomic layer of graphite \cite{grp}, have a much smaller speed of propagating sound waves. 
Thus, if there is any hope of realistically modeling such systems in the context AdS/CFT 
duality, one needs to understand how to 
reduce the speed of sound in holographic relativistic plasma. In \cite{sound} it was argued that
sound waves in 3+1 dimensions are coupled to the magnetic 
variables of the fluid since the magnetization oscillates with the fluid density.
As a result, dispersion relation of the sound waves is affected by the 
external magnetic fields. We would like to understand 
here whether such an effect persists in strongly coupled (2+1) dimensional magnetic fluids which admit 
a dual holographic realization.
    
In this paper, we explore 
viscous hydrodynamics of strongly coupled $M2$-brane plasma in external magnetic field. 
In the next section we discuss magneto-hyd\-ro\-dy\-na\-mics from the gauge theory 
perspective, emphasizing the necessity to appropriately scale 
the external magnetic field in the hydrodynamic limit. 
We point out that decoupling of the shear and sound modes in magnetic fluids requires vanishing of the 
equilibrium charge density (or the corresponding chemical potential). We derive dispersion relation for the 
propagation of the sound waves, explicitly demonstrating its sensitivity to the background field, similar to 
what was observed in \cite{sound}. 
In particular, we show that we can reduce the speed of sound by appropriately 
turning on a magnetic field.
In section 3 we analyze magneto-hydrodynamics of dyonic black holes in 
$AdS_4\times S^7$ supergravity backgrounds of M-theory. 
The latter realizes a holographic dual to strongly 
coupled $M2$-brane magnetized plasma.  
We study quasinormal modes of these dyonic black holes, and demonstrate,
in parallel with the field-theoretic analysis, that decoupling of the shear and the sound modes occur only 
for black holes with zero electric charge. We compute dispersion 
relation of sound quasinormal modes, and interpret the 
results within the framework of HKMS magneto-hydrodynamics. Finally, we conclude in section 4.


\section{Gauge Theory Magneto-Hydrodynamics}


\subsection{Equations of Motion and Conformal Invariance}


In this paper, we are interested in hydrodynamic 
properties of the (2+1)-dimensional theory living on a 
large number of $M2$-branes. One can view this theory as 
the three-dimensional maximally supersymmetric gauge theory 
near the infrared fixed point. It also admits a holographic 
description as M-theory on the manifold 
$AdS_4 \times S^7$~\cite{m9711}. 
The state corresponding to the thermal equilibrium is
described by the black brane in $AdS_4$~\cite{Wittenfirst}. 
The most rigorous way to understand the field theory equations
of motion is to use this holographic duality. For a review of
the near-boundary holographic analysis 
of asymptotically $AdS$ space-times see~\cite{Skenderis}. 

Let $g_{M N}$ and $A_{M}$, $M, N=0, \dots, 3$ be the metric 
and the gauge field in $AdS_4$.\footnote{For purposes of the paper, we
will specialize to the case of the four-dimensional $AdS$ space 
with three-dimensional boundary and restrict ourselves to the  
the metric and gauge bulk fields. Of course, this analysis can be made more general. 
See~\cite{Skenderis} and references therein for details.}
To construct the dual field theory on the boundary, we have to solve 
the equations of motion for $g_{M N}$ and $A_{M}$
with appropriately defined
boundary conditions $g_{M N} \to g_{\mu \nu}^{(0)}$ and $A_{M} \to A_{\mu}^{(0)}$, 
$\mu, \nu =0, 1, 2$.
According to the AdS/CFT
dictionary~\cite{Gubser, Wittenfirst}, the boundary correlation 
functions are encoded in the {\it renormalized} action
$S_{ren}[g_{\mu \nu}^{(0)}, A_{\mu}^{(0)}]$. In constructing $S_{ren}[g_{\mu \nu}^{(0)}, A_{\mu}^{(0)}]$
one can perform the integral over the $AdS$ radial coordinate and it becomes 
a functional on the boundary. This procedure of constructing the 
renormalized boundary action is known by the name ``holographic renormalization''. 
Then, the boundary stress-energy tensor and the current are defined as
\begin{equation}
\delta S_{ren}[g_{\mu \nu}^{(0)}, A_{\mu}^{(0)}]=\int d^3x \sqrt{-g}(\frac{1}{2}
\langle T_{\mu \nu}\rangle \delta g^{(0) \mu \nu}+
\langle J^{\mu}\rangle \delta A_{\mu}^{(0)})\,.
\eqlabel{2.1}
\end{equation}
The field theory equations of motion become the consequences of the symmetries 
of the action $S_{ren}$. Since $S_{ren}$ is invariant under diffeomorphisms
and gauge transformations it follows that we have the boundary
conservation equations of the form 
\bea
\nabla_{\nu}T^{\mu \nu}&=&F^{\mu \nu}J_{\nu} \nonumber\,, \nonumber \\
\nabla_{\mu}J^{\mu}&=&0\,, 
\label{cfteom}
\eea
where, to simplify notation, we have removed the brackets $\langle \rangle$.
To derive eqs.~\eqref{cfteom} one has to assume that the diffeomorphisms and gauge transformations 
act non-trivially on the boundary. However, in the bulk there is a special type of 
diffeomorphism which does not transform the coordinates $x^{\mu}$ on the boundary
and whose boundary effect is to Weyl rescale 
the metric $g_{\mu \nu}^{(0)}$~\cite{Imbimbo}. 
Such diffeomorphism can also act non-trivially on the 
boundary gauge potential~\cite{Freedman}.
In Appendix A  we show that the induced transformation on the boundary 
gauge field is trivial for a massless bulk gauge field. 
Therefore, under such diffeomorphism, the boundary fields transform as
\begin{equation}
\delta g^{(0) \mu \nu}=-2 \sigma(x)g^{(0) \mu \nu}\,, \quad 
\delta A_{\mu}^{(0)} =0 \,,
\label{weylxform}
\end{equation}
where $\sigma(x)$ is the transformation parameter.  
See eq.~\eqref{diff} in Appendix A.
The renormalized action $S_{ren}$ has to be invariant under all 
the bulk diffeomorphisms.
The invariance under these transformations implies 
that\footnote{The right hand side of this equation has to be supplemented 
by the conformal anomaly which can also be computed 
holographically~\cite{Henn}. However, it is not relevant for 
our discussion since the conformal anomaly vanishes in odd dimensions. 
In the case of four-dimensional hydrodynamics it is relevant at fourth 
order as pointed out in~\cite{Baier}.}
\begin{equation}
T_{\mu}^{{\ }\mu}=0\,.
\eqlabel{2.9}
\end{equation}
Therefore, conformal invariance is unbroken by the presence of the background gauge field.
The unbroken conformal invariance can also be understood intuitively
if one recalls that
the fact that the gauge field goes to a finite $x^\mu$ dependent piece $A^{(0)}_\mu(x)$ on the boundary (see Appendix A), means that 
in field theory it represents a marginal deformation~\cite{Wittenfirst}.

In this paper, we consider hydrodynamics in the presence of the 
net charge density $\rho$ and the background magnetic field $B$.
For simplicity, we will 
drop the label ``$(0)$" from the boundary fields in what follows
and denote the bulk and the boundary fields by the same letter. 
However, from the context it will be clear whether the corresponding field 
belongs to the bulk or to the boundary.


\subsection{The First Order Hydrodynamics}


In this subsection, we study hydrodynamic perturbations 
in the presence of the charge density $\rho$ and the magnetic 
field $B$. For our purposes, we can take the boundary metric to be flat
and the equations of motion become
\begin{eqnarray}
&&\partial_{\nu}T^{\mu \nu}=F^{\mu \nu}J_{\nu}\,, \nonumber \\
&&\partial_{\mu}J^{\mu}=0\,,
\label{2.10}
\end{eqnarray}
where $F^{\mu \nu}$ is the field strength corresponding to the 
background electromagnetic field. 
One can do the standard decomposition of the stress tensor,
\begin{equation}
T^{\mu \nu} =\epsilon u^{\mu}u^{\nu}+P \Delta^{\mu \nu} +\Pi^{\mu \nu}\,, 
\eqlabel{2.11}
\end{equation}
where 
\begin{equation}
\Delta^{\mu \nu}=\eta^{\mu \nu}+u^{\mu}u^{\nu}\,,\;\;\; \Pi^{\mu}_{\;\;\nu} u^\nu = 0\;,
\eqlabel{2.12}
\end{equation}
and $\epsilon$ and $P$ are the energy density and the pressure respectively.  In writing eq. (\ref{2.11}) we are working in the so-called ``energy frame", and the four velocity of the fluid $u^\mu$ is {\it defined} by the eigenvalue equation  $T^\mu_\nu u^\nu = - \epsilon u^\mu$. 
It was argued in~\cite{Kovtun} that in the presence of the magnetic field 
the pressure $P$ differs from the usual thermodynamic pressure $p$ by the term
$-M B$, where $M$ is the magnetization.
The dissipative term $\Pi^{\mu \nu}$ is given by
\begin{equation}
\Pi^{\mu \nu}=-\eta \sigma^{\mu \nu} -\zeta \Delta^{\mu \nu} (\partial_{\alpha} u^{\alpha})\,, 
\eqlabel{2.14}
\end{equation}
where 
\begin{equation}
\sigma^{\mu \nu}=\Delta^{\mu \alpha} \Delta^{\nu \beta}
(\partial_{\alpha}u_{\beta}+\partial_{\beta}\partial_{\alpha})
-\Delta^{\mu \nu} (\partial_{\gamma}u^{\gamma})\,,
\eqlabel{2.15}
\end{equation}
and $\eta$ and $\zeta$ are the shear and bulk viscosity. Note that $\Pi^{\mu \nu}$ is, by definition, zero at local equilibrium. 

Now we consider the current $J^{\mu}$. It is given by 
\begin{equation}
J^{\mu}=\rho u^{\mu} +\nu^{\mu}\,, 
\eqlabel{2.16}
\end{equation}
where $\nu^{\mu}$ is the dissipative part satisfying $u^{\mu}\nu_{\mu}=0$. 
The expression for it can be obtained from the argument
that the entropy production has to be positive. It was done in~\cite{Kovtun}
and the result is
\begin{equation}
\nu^{\mu}=\sigma_{Q}\Delta^{\mu \nu}(-\partial_{\nu}\mu
+F_{\nu \alpha}u^{\alpha}+\frac{\mu}{T}\partial_{\nu}T)\,. 
\eqlabel{2.17}
\end{equation}
In this expression, $T$ is the temperature, $\mu$ is the chemical potential, 
$F_{\mu \nu}$ is the background field strength and $\sigma_Q$ is the DC conductivity coefficient. 
In the case under consideration
\begin{equation}
F_{0 i}=0\,, \quad i=1, 2\,, \qquad F_{i j}=\epsilon_{i j} B\,. 
\eqlabel{2.17.1}
\end{equation}

We would like to study fluctuations around the equilibrium state in which 
\begin{equation}
u^{\mu}=(1, 0, 0)\,, \quad T={\rm const.}\,, \quad \mu={\rm const.}\,.
\eqlabel{2.18}
\end{equation}
As an independent set of variables we will choose the two components 
of the velocity $\delta u_1\equiv \delta u_x$, $\delta u_2\equiv \delta u_y$
as well as $\delta T$ and $\delta \mu$. As usual, all perturbations 
are of the plane-wave form $exp(-i \omega t+i q y)$. We find that the relevant 
fluctuations of $T^{\mu \nu}$ are
\begin{equation}
\begin{split}
& \delta T^{t t} =\delta \epsilon = 
\left(\frac{\partial \epsilon}{\partial \mu}\right)_{T}
\delta \mu 
+\left(\frac{\partial \epsilon}{\partial T}\right)_{\mu}
\delta T\,,  \\
& \delta T^{t i}=(\e +P)\delta u^i\,, \\
&\delta T^{x y}=-\eta \partial_y \delta u_x\,,  \\
&\delta T^{y y}=\delta P- (\eta +\zeta)\partial_y \delta u_y=
\left( \frac{\partial P}{\partial \mu}\right)_{T}
\delta \mu+
\left( \frac{\partial P}{\partial T}\right)_{\mu}
\delta T 
-(\eta+\zeta)\partial_y \delta u_y\,. 
\end{split}
\label{2.19}
\end{equation}
Note that since $P$ is different from the thermodynamic pressure, 
$\left( \frac{\partial P}{\partial \mu}\right)_{T}$ and 
$\left( \frac{\partial P}{\partial T}\right)_{\mu}$ do not coincide
with the charge density $\rho$ and the entropy density $s$. 
The equalities, however, hold when we set $B=0$. Similarly, we obtain the 
following fluctuations of the current
\begin{eqnarray}
&&
\delta J^t =\delta \rho= 
\left(\frac{\partial \rho}{\partial \mu}\right)_{T}
\delta \mu +
\left(\frac{\partial \rho}{\partial T}\right)_{\mu}
\delta T\,, \nonumber \\
&& \delta J^x =\rho \delta u_x +\sigma_QB \delta u_y\,, \nonumber \\
&& \delta J^y= \rho \delta u_y+
\sigma_Q (-\partial_y \delta\mu+ \frac{\mu}{T}\partial_y \delta T- B\delta u_x)\,. 
\label{2.20}
\end{eqnarray}
Substituting these expressions into equations of motion~\eqref{2.10}
and performing a Fourier transformation we get the following system of equations
\begin{equation}
\begin{split}
0=&\omega 
\left(\left(\frac{\partial \epsilon}{\partial \mu}\right)_{T}
\delta \mu+\left(\frac{\partial \epsilon}{\partial T}\right)_{\mu}\delta T\right)-
q (\e+P) \delta u_y\,,\\
0=&\omega (\epsilon+P)\delta u_y -q
\left( \left(\frac{\partial P}{\partial \mu}\right)_T \delta \mu
+\left(\frac{\partial P}{\partial T}\right)_{\mu}
\delta T\right)+i q^2(\eta +\zeta)\delta u_y
+i \sigma_Q B^2 \delta u_y\,, \\
&+i B \rho \delta u_x\,,\\
0=&\omega (\epsilon+P) \delta u_x - q B \sigma_Q (\delta \mu -\frac{\mu}{T} \delta T)
-i B\rho \delta u_y+ i \sigma_Q B^2 \delta u_x +i q^2 \eta \delta u_x\,,\\
0=&\omega 
\left(\left(\frac{\partial \rho}{\partial \mu}\right)_{T}\delta \mu+
\left(\frac{\partial \rho}{\partial T}\right)_{\mu}\delta T\right)
-q \rho \delta u_y +q \sigma_Q B \delta u_x+ i q^2\sigma_Q (\delta \mu -\frac{\mu}{T}\delta T)\,.
\end{split}
\eqlabel{2.21}
\end{equation}

Our aim is to understand how sound and shear modes are modified in the 
presence of the charge density and the magnetic field.
However, eqs.~\eqref{2.21} are all coupled to each other and it does not 
seem to be meaningful to ask what happens to the sound and shear modes 
separately. On the other hand, there is a regime in which these equations
simplify and decouple into the two independent pairs. 
First, we will consider hydrodynamics with no charge density and no chemical potential 
\begin{equation}
\rho=0\,, \qquad \mu=0\,. 
\eqlabel{2.22}
\end{equation}
In addition, motivated by $M2$-brane magneto-hydrodynamics,
we will set
\begin{equation}
\left(\frac{\partial \rho}{\partial T}\right)_{\mu}=0\,, \quad 
\left(\frac{\partial \epsilon}{\partial \mu}\right)_{T}=0\,, \quad 
\left(\frac{\partial \rho}{\partial \mu}\right)_{T} \neq 0\,, \quad 
\left(\frac{\partial \epsilon}{\partial T}\right)_{\mu} \neq0\,. 
\eqlabel{2.23}
\end{equation}
Conditions~\eqref{2.23} are also satisfied at $\rho=0$, $\mu=0$ on the supergravity side
which will be studied in the next section. 
Then, equations~\eqref{2.21} get separated into the
two decoupled pairs. The first pair reads
\begin{eqnarray}
&&\omega \left(\frac{\partial \epsilon}{\partial T}\right)_{\mu}\delta T -
q (\epsilon+P)\delta u_y=0\,, \nonumber \\
&& \omega  (\epsilon+P)\delta u_y-
q \left(\frac{\partial P}{\partial T}\right)_{\mu} \delta T+
i q^2 (\eta +\zeta)\delta u_y +i \sigma_Q B^2 \delta u_y=0\,. 
\label{2.24}
\end{eqnarray}
If we set $B=0$, these two equations describe the sound mode with dispersion relation 
\begin{equation}
\omega=\pm c_s q -i q^2 \frac{\eta +\zeta}{\epsilon +P}\,, 
\eqlabel{2.25}
\end{equation}
where the speed of sound is defined, as usual, as $c_s^2=\frac{\partial P}{\partial \epsilon}$. 
We will refer to this pair of equation as to the sound mode equations. 
The second decoupled pair of equations becomes
\begin{eqnarray}
&&\omega 
(\e+P) \delta u_x -q B \sigma_Q \delta\mu +i \sigma_Q B^2 \delta u_x
+ i q^2 \eta \delta u_x=0\,, \nonumber \\
&& \omega \left(\frac{\partial \rho}{\partial \mu}\right)_T\delta \mu +
q \sigma_Q B \delta u_x +
i q^2 \sigma_Q \delta \mu=0\,.
\label{2.26}
\end{eqnarray}
In the absence of the magnetic field these two equations describe a
shear perturbation $\delta u_x$ with dispersion relation 
\begin{equation}
\omega =-i q^2 \frac{\eta}{\epsilon+P}\,,
\eqlabel{2.27}
\end{equation}
and a diffusive mode $\delta \mu$ with dispersion relation 
\begin{equation}
\omega =-i q^2 \frac{\sigma_Q}{\left(\frac{\partial\rho}{\partial \mu}\right)_T}\,. 
\eqlabel{2.28}
\end{equation}
We will refer to these equations as to the shear mode equations.

In the presence of the magnetic field all these solutions~\eqref{2.25}, \eqref{2.27} 
and~\eqref{2.28} disappear. From the sound mode equations we obtain one constant solution
\begin{equation}
\omega=-i \frac{\sigma_Q B^2}{\epsilon+P} +{\cal O}(q^2)\,,
\eqlabel{2.29}
\end{equation}
and a diffusive mode
\begin{equation}
\omega=-i q^2 c_s^2 \frac{\epsilon+P}{\sigma_Q B^2} +{\cal O}(q^4)\,. 
\eqlabel{2.30}
\end{equation}
From the shear mode equations we also obtain a constant mode~\eqref{2.29}
and a subdiffusive mode
\begin{equation}
\omega=-i q^4 \frac{\eta}{B^2 \left(\frac{\partial \rho}{\partial \mu}\right)_T}+{\cal O}(q^6)\,. 
\eqlabel{2.31}
\end{equation}
In this analysis it has been assumed that the magnetic field $B$ is held 
fixed in the hydrodynamic limit. 

However, it is not clear whether the solutions~\eqref{2.30} and~\eqref{2.31}
can be trusted.
The first hint that the regime of constant $B$ in the 
hydrodynamic limit might not be well-defined comes from inspecting the $B$-dependent 
term in eq.~\eqref{2.17}. The conductivity coefficient $\sigma_Q$ is of order 
the free mean path ${\ell}$. Therefore, at $B=0$ each term in~\eqref{2.17}  
is of order $\frac{\ell}{L}$ where $L$ is the scale over which the derivatives vary. 
However, the term with $F_{\nu \alpha}$ at constant $B$ is not of this order
since this term does not contain derivatives. This means that this term is not small 
in the hydrodynamic limit $\frac{\ell}{L}\ll 1$. It also follows that the limit 
of small $B$ does not commute with the hydrodynamic limit. 
In other words, $B$ cannot be thought of as being a small perturbation and 
one can worry that 
the hydrodynamic analysis in this case is unstable under higher order corrections. 
Let us now present a more quantitative reason why the solutions given above 
might not be reliable. From equations~\eqref{2.24} it follows that 
if $\omega \sim q^2$ we obtain 
\begin{equation}
\frac{\delta T}{\delta u_y}\sim \frac{1}{q}\,.
\eqlabel{2.32}
\end{equation}
Thus, assuming that the amplitude $\delta u_y$ is fixed and of order unity 
in the hydrodynamic limit, we find that the amplitude of $\delta T$ is infinitely large.
This means that the terms which are naively of higher order 
because they are suppressed by higher power of $\omega$ and $q$ can, in fact, 
modify hydrodynamics at lower order because of the large amplitude.

Hence, it is more natural to study magnetic fields which vanish in the hydrodynamic limit.
That is, we consider $B$ which scales as 
\begin{equation}
B=b q^p\,, \quad p>0\,,
\eqlabel{2.33}
\end{equation}
with $b$ held fixed. 
An interesting observation is that 
choosing different values of $p$ we can probe hydrodynamics
in different regimes. 
In this paper, we will concentrate on the sound waves for $p=1$ and $p=1/2$. 
Below we will present the field theory results and
in the next section, we will study the holographic dual description.

Similar analysis can also be performed for the shear modes.
We will not do it in the present paper. 


\subsection{Sound Waves in Magnetic Field}


If the magnetic field vanishes in the hydrodynamic limit, 
to leading order in $q$
we can consider the various transport coefficients 
and susceptibilities evaluated at $B=0$. This, of course, is consistent
if we are interested in the first order hydrodynamics. 
If one wishes to go to the second (or higher) order
one has to keep in mind that there will be corrections 
not only from the term which are of higher order in derivatives but also
from the possible $B=b q^p$-dependence of the transport coefficients.
Thus, the shear viscosity $\eta$ and the conductivity $\sigma_Q$ can be taken to be
equal to their values at $B=0$.  
Since, as shown in subsection 2.1, our theory is conformally invariant
the bulk viscosity $\zeta$ vanishes.
Furthermore, the pressure $P$ in this case becomes the usual
thermodynamic pressure $p$. Therefore, we have 
\begin{equation}
\left(\frac{\partial P}{\partial T}\right)_{\mu}=
\left(\frac{\partial p}{\partial T}\right)_{\mu}=s\,. 
\eqlabel{2.34}
\end{equation}
In addition, we have another well-known relation 
\begin{equation}
\delta \epsilon =c_s^2 \delta P\,, 
\eqlabel{2.35}
\end{equation}
where $c_s^2=1/2$. 
With these simplifications, we have the following sound mode equations
\begin{eqnarray}
&& \omega \delta \epsilon -q (\epsilon +P)\delta u_y=0\,, \nonumber \\
&& \omega (\epsilon +P) \delta u_y -\frac{1}{2}q \delta \epsilon +i q^2 \eta \delta u_y
+i \sigma_Q B^2 \delta u_y=0\,. 
\label{2.36}
\end{eqnarray}
We would like to study solutions to these equations when $B$ scales as
$b q$ and $b q^{1/2}$. 


\subsubsection{The Regime $B=b q$}


Substituting $B=b q$ into eqs.~\eqref{2.36} we obtain
\begin{eqnarray}
&& \omega \delta \epsilon -q (\epsilon +P)\delta u_y=0\,, \nonumber \\
&& \omega (\epsilon +P) \delta u_y -\frac{1}{2}q \delta \epsilon +
i q^2 (\eta+ \sigma_Q b^2) \delta u_y=0\,. 
\label{2.37}
\end{eqnarray}
We see that the effect of the magnetic field is to shift the shear viscosity 
by the amount $\sigma_Q b^2$. This means that the modified dispersion relation is 
\begin{equation}
\omega= \pm \frac{1}{\sqrt{2}}q -\frac{i q^2}{2} \frac{\eta +\sigma_Q b^2}{\epsilon +P}\,. 
\eqlabel{2.38}
\end{equation}
We obtain a sound wave whose speed is $1/\sqrt{2}$ of the speed of light
and with modified attenuation. 


\subsubsection{The Regime  $B=b q^{1/2}$}


Substituting $B=b q^{1/2}$ into eq.~\eqref{2.36} we obtain
\begin{eqnarray}
&&\omega \delta \epsilon- q(\epsilon+P)\delta u_y=0\,, \nonumber \\
&& \omega (\epsilon +P) \delta u_y -\frac{1}{2}q \delta \epsilon +
i q^2 \eta \delta u_y  +i \sigma_Q b^2 q  \delta u_y  = 0\,,
\label{2.39}
\end{eqnarray}
and the corresponding characteristic equation becomes
\begin{equation}
\omega^2 (\epsilon+P) +i \omega (\eta q^2 + \sigma_Q b^2 q)-\frac{1}{2}q^2 (\epsilon+P)=0\,. 
\eqlabel{2.40}
\end{equation}
Note that if we take $\sigma_Q$ and $(\epsilon+P)$ to be equal to their values at 
$B=0$ we cannot trust the $\omega q^2$ term in this equation since it will be modified 
due to $B^2$ dependence of $\sigma_Q$ and $(\epsilon+P)$. Then the solution becomes
the following sound wave
\begin{eqnarray}
\omega_{1, 2} & = &-i \frac{\sigma_Q b^2}{2 (\epsilon +P)}q\pm 
\frac{q}{\sqrt{2}}\sqrt{1-\frac{\sigma_Q^2 b^4}{2 (\epsilon+P)^2}} \nonumber \\
 & = & -i \frac{\sigma_Q b^2}{2 (\epsilon +P)}q \pm \frac{q}{\sqrt{2}}
\left( 1-\frac{\sigma_Q^2 b^4}{4 (\epsilon+P)^2} +{\cal O}(b^6)\right)\,.
\label{2.41}
\end{eqnarray}
Note that the speed of this sound wave is different from 
$1/\sqrt{2}$ and is decreased in presence of the magnetic field. 

Now let us compute
corrections of order $q^2$
to the dispersion relation. Going back to eqs.~\eqref{2.24}
we see that to obtain all necessary terms to the given order
we need to expand $(\epsilon+P), \frac{\partial \epsilon}{\partial T}, 
\frac{\partial P}{\partial T}, \sigma_Q$ to next-to-leading order in 
$B^2=b^2 q$. Note that $(\eta+\zeta)$ already multiplies $q^2$ and, hence, can 
be taken in the limit of the zero magnetic field. Furthermore, from subsection 2.1
we know that the theory is conformally 
invariant. Hence,
%
\begin{equation}
P=\frac{\epsilon}{2}\,,
\eqlabel{2.42}
\end{equation}
where, as we have explained before, $P=p-M B$, with $p$ being the thermodynamic 
pressure and $M$ being the magnetization. 
Then it follows that 
eq.~\eqref{2.40} still holds but we have to expand $(\epsilon+P)$ and 
$\sigma_Q$ to next-to-leading order in $B^2=b^2 q$. Let us denote 
\begin{eqnarray}
&& \epsilon+P=E_0 +q b^2 E_1\,, \nonumber \\
&& \sigma_Q =\sigma_0+ q b^2 \sigma_1\,. 
\label{2.43}
\end{eqnarray}
For simplicity, we will work to order $b^4$. Then we obtain the following 
solutions to~\eqref{2.40}
\begin{eqnarray}
\omega_{1, 2} & = & -\frac{i q}{2 E_0}
\left( \frac{\sigma_0 b^2}{E_0}+ \eta q + \left(\sigma_1 -\frac{E_1}{E_0}\right)b^4 q\right)
\nonumber \\
 & \pm & \frac{q}{\sqrt{2}}\left(1 -\frac{\sigma_0^2 b^4}{4 E_0^2} -
\frac{\sigma_0 \eta b^2}{2 E_0^2}q \right)\,.
\label{2.44}
\end{eqnarray}

This finishes our field theory consideration. Now we are going to
move to the supergravity side. Our aim will be to reproduce the solutions
discussed in this section.


\section{Supergravity Magneto-Hydrodynamics}


\subsection{Effective Action and Dyonic Black Hole Geometry}


The effective four-dimensional bulk action describing supergravity 
dual to $M2$-brane plasma in the
external field is given by~\cite{HK}\footnote{For simplicity, we set the radius 
of $AdS_4$ to unity.}
\begin{equation}
S_4=\frac{1}{g^2} \int dx^4 \sqrt{-g}\left[-\frac 14  R +\frac 14 F_{M N}F^{M N}
-\frac 32\right]\,,
\eqlabel{4action}
\end{equation} 
where 
\begin{equation}
\frac{1}{g^2}=\frac{\sqrt{2}N^{3/2}}{6\pi}\,,
\eqlabel{defa}
\end{equation}
and $g$ is the bulk coupling constant. 
From eq.~\eqref{4action} we obtain the following equations of motion
\begin{equation}
\begin{split}
&R_{M N}=2 F_{M L}F_{N}^{\ L}-\frac 12 g_{M N}F_{L P}
F^{L P}-3 g_{M N}\,,\\
&\nabla_M F^{M N}=0\,.
\end{split}
\eqlabel{eom}
\end{equation}
According to the AdS/CFT dictionary, the equilibrium 
state of magneto-hydrodynamics is described by dyonic black hole geometry
whose Hawking temperature is identified with the plasma temperature
on the field theory side. 
A dyonic black hole in $AdS_4$ with planar horizon is 
given by the following solution to eqs.~\eqref{eom}~\cite{HK}
\begin{equation}
\begin{split}
ds^2_4\equiv& -c_1(r)^2\ dt^2+c_2(r)^2\ \left[dx^2+dy^2\right]+c_3(r)^2\ dr^2\\
=&\frac {\a^2}{r^2}\left[-f(r) dt^2+dx^2+dy^2\right]+\frac {1}{r^2}
\frac{dr^2}{f(r)}\,,\\
F=&h\a^2 dx\wedge dy+q \a dr\wedge dt\,,\\
 f(r)=&1+(h^2+q^2)r^4
-(1+h^2+q^2) r^3\,,
\end{split}
\eqlabel{metric}
\end{equation}
where $h,q,\a$ are constants related to the field theory quantities as 
follows~\cite{HK}
\begin{equation}
B=h\a^2\,,\qquad \mu=-q\a\,,\qquad \frac{4\pi T}{\a}=3-\frac{\mu^2}{\a^2}-\frac{B^2}{\a^4}\,,
\eqlabel{hqa}
\end{equation} 
where $B$ is the external magnetic field of the equilibrium $M2$-brane plasma, $\mu$ is its 
chemical potential, and $T$ is the plasma temperature.

Let us now review thermodynamics of dyonic black holes. 
It has been studied extensively in~\cite{HK}
to which we refer for additional details.
In the grand canonical ensemble
the thermodynamic potential is given by
\begin{equation}
\Omega =-V_2\ p= V_2\ \frac{1}{g^2}   
\frac{\a^3}{4} \left(-1-\frac{\mu^2}{\a^2}+3\frac{B^2}{\a^4}\right)\,,
\eqlabel{th1}
\end{equation}
where $V_2$ is area of the $(x, y)$-plane and
$p$ is the thermodynamic pressure.
Just like on the field theory side, as the independent 
quantities we consider $\alpha$ related to the temperature by eq.~\eqref{hqa}, 
the chemical potential $\mu$ and the magnetic field $B$. In terms of these variables 
we can compute the energy $\epsilon$, the entropy $s$ and the electric charge $\rho$
per unit area. 
One obtains~\cite{HK}
\begin{equation}
\e=\frac{1}{g^2} \ \frac{\a^3}{2}\left(1+\frac{\mu^2}{\a^2}+\frac{B^2}{\a^4}\right)\,.
\eqlabel{th2}
\end{equation}
Just like on the field theory side, 
it coincides with the temporal component $\langle T_{00}\rangle$  
of the stress-energy tensor. Furthermore,
the entropy density is given by
\begin{equation}
s=\frac{\pi}{g^2} \ \a^2\,.
\eqlabel{th3}
\end{equation}
Finally, 
the charge density is
\begin{equation}
\r=\frac{1}{g^2} \ \a\mu\,.
\eqlabel{th4}
\end{equation}
In addition, we introduce the magnetization per unit area
\begin{equation}
M=-\frac{1}{V_2}\left(\frac{\partial \Omega}{\partial B}\right)_{T, \mu}=
-\frac{1}{g^2} \ \frac{B}{\a}\,.
\eqlabel{th5}
\end{equation}
In the presence of the magnetic field, the thermodynamic pressure $p$ 
is different from $\langle T_{x x}\rangle$ by the term proportional to the magnetization. 
Just like on the field theory side, we introduce the pressure $P$ as $\langle T_{x x}\rangle$
which equals to
\begin{equation}
P=p- M B\,. 
\eqlabel{3.10.1}
\end{equation}
It is easy to check that 
\begin{equation}
P=\frac{\epsilon}{2}\,.
\eqlabel{th6}
\end{equation}
Notice that the trace of the stress-energy tensor  
\begin{equation}
\langle T_{\nu}^{\ \nu}\rangle=0
\eqlabel{trace}
\end{equation}
vanishes, implying unbroken scale invariance.

In parallel to the field theory discussion in the previous section 
we study sound quasinormal modes of the {\it magnetically} charged black hole, \ie, we set 
\begin{equation}
q=0\qquad \Longleftrightarrow\qquad \mu=0\,.
\end{equation}
Then
from eqs.~\eqref{th1}-\eqref{th6} we obtain
\begin{equation}
\begin{split}
&\left(\frac{\del\e}{\del T}\right)_{\mu}
=\frac{1}{g^2}\ \frac{2\a^2\pi(3\a^4-B^2)}{3(\a^4+B^2)}\,,\qquad 
\left( \frac{\del\e}{\del\mu} \right)_{T}=0\,,
\\
&\left(\frac{\del\r}{\del T}\right)_{\mu}=0\, ,
\qquad \left(\frac{\del\r}{\del \mu}\right)_{T}=\frac{1}{g^2}\ \a\,.
\end{split}
\eqlabel{dth}
\end{equation}
This is in exact agreement with our field theory conditions~\eqref{2.23}.


\subsection{Fluctuations}


Now we study fluctuations in the background geometry
\begin{equation}
\begin{split}
g_{M N}&\to g_{M N}+h_{M N}\,,\\
A_M&\to A_M+a_M\,,
\end{split}
\eqlabel{fluctuations}
\end{equation}
where $g_{M N}$  and $A_M$ ($dA=F$) are the black brane 
background configuration \eqref{metric},
and $\{h_{M N}, a_{M}\}$ are the fluctuations. 
To proceed, it is convenient
to choose the gauge 
\begin{equation}
h_{tr}=h_{xr}=h_{yr}=h_{rr}=0\,,\qquad a_r=0\,.
\eqlabel{gaugec}
\end{equation}
To be consistent with the field theory side, we will take  
all the fluctuations to depend only on $(t,y,r)$,\ \ie, we have a $\zet_2$ parity symmetry along the 
$x$-axis. Strictly speaking, for this parity to be a symmetry the reflection of the $x$ coordinate 
must be accompanied by the change of the $B$-field orientation:
\begin{equation}
\zet_2:\qquad x\to -x\qquad \&\qquad h\to -h\,.
\eqlabel{zet2}
\end{equation}

At a linearized level, and for the vanishing chemical potential $\mu=0$,  
we find that the following sets of fluctuations decouple from each other
\begin{equation}
\begin{split}
\zet_2-{\rm even}:\qquad &\{h_{tt},h_{ty},h_{xx},h_{yy};a_x\}\,,\\
\zet_2-{\rm odd}:\qquad &\{h_{tx},h_{xy};a_t,a_y\}\,.\\
\end{split}
\eqlabel{n1}
\end{equation}
Notice that naively the gauge potential fluctuation  $a_x$ is parity-odd, while the other 
two components are parity-even. This is misleading, as it will turn out that $a_{M}\propto h$, and 
thus $a_x$ is parity-even, while $\{a_t,a_y\}$ are parity-odd. 
The first set of fluctuations is a  holographic dual to the sound waves in the $M2$-brane  
plasma in the 
external magnetic field,  
which is of interest here.
The second set describes the shear and diffusive modes.
Note that if the electric charge of the black hole $q$ and, hence, 
the chemical potential is not zero,
$a_M$ is a linear combination of the terms some of which are 
proportional to $h$ and some are proportional to $q$. Therefore, in this case
$a_M$ does not have a definite sign under parity~\eqref{zet2}. As the result, the two sets of
fluctuations in eq.~\eqref{n1} no longer decouple. To say it differently, 
if both $h$ and $q$ are non-zero,
the electromagnetic background~\eqref{metric} is not an eigenstate 
of the spacial parity $x\to -x$. However, the decoupling of the sound and shear 
fluctuations requires that it be an eigenstate. 
This is in complete agreement with our field theory analysis. 
Eqs.~\eqref{2.21} decouple into the two separate pairs of equations 
only if we set $\mu=0$, $\rho=0$. 
  
Let us introduce
\begin{equation}
\begin{split}
h_{tt}=&c_1(r)^2\ \hh_{tt}=e^{-i\w t+iq y}\ c_1(r)^2\  H_{tt}\,,\\
h_{ty}=&c_2(r)^2\ \hh_{ty}=e^{-i\w t+iq y}\ c_2(r)^2\  H_{ty}\,,\\
h_{xx}=&c_2(r)^2\ \hh_{xx}=e^{-i\w t+iq y}\ c_2(r)^2\  H_{xx}\,,\\
h_{yy}=&c_2(r)^2\ \hh_{yy}=e^{-i\w t+iq y}\ c_2(r)^2\  H_{yy}\,,\\
a_x=&i e^{-i\w t+iq y}\ \ha_x\,,
\end{split}
\eqlabel{rescale}
\end{equation} 
where $\{H_{tt},H_{ty},H_{xx},H_{yy},\ha_x\}$ are functions of the radial coordinate only
and $c_1(r)$ and $c_2(r)$ are defined in eq.~\eqref{metric}.
Expanding at a linearized level eqs.~\eqref{eom} 
using
eqs.~\eqref{fluctuations} and eqs.~\eqref{rescale} we find
the following coupled system of ODE's
\begin{equation}
\begin{split}
0=&H_{tt}''+H_{tt}'\ \left[\ln\frac{c_1^2c_2}{c_3}\right]'+
\frac 12\left[H_{xx}+H_{yy}\right]'\ \left[\ln \frac{c_2}{c_1}\right]'
-\frac{c_3^2}{2c_1^2}\biggl(q^2\frac{c_1^2}{c_2^2}\ (H_{tt}+H_{xx})\\
&+\w^2\ (H_{xx}+H_{yy})+2\w q\ H_{ty}\biggr)
-3\ \frac{c_3^2}{c_2^4}\ h^2\a^4\ (H_{xx}+H_{yy})+6\ \frac{c_3^2}{c_2^4}\ h\a^2 q\ \ha_x\,,
\end{split}
\eqlabel{fl1}
\end{equation}
\begin{equation}
\begin{split}
0=&H_{ty}''+H_{ty}'\ \left[\ln\frac{c_2^4}{c_1c_3}\right]'
+\frac{c_3^2}{c_2^2}\ \w q\ H_{xx}-4 \frac{c_3^2}{c_2^4}\ h\a^2 \left(h\a^2\ H_{ty}+\w\ \ha_x\right)\,,
\end{split}
\eqlabel{fl2}
\end{equation}
\begin{equation}
\begin{split}
0=&H_{xx}''+\frac 12 H_{xx}'\ \left[\ln\frac{c_1^5c_2}{c_3^2}\right]'+
\frac 12\  H_{yy}'\ \left[\ln \frac {c_2}{c_1}\right]'
+\frac{c_3^2}{2c_1^2}\biggl(\w^2(H_{xx}-H_{yy})-q^2\frac{c_1^2}{c_2^2}( H_{tt}+H_{xx})\\
&-2\w q H_{ty}\biggr)
-\frac{c_3^2}{c_2^4}\ h^2\a^4\ (H_{xx}+H_{yy})+2\ \frac{c_3^2}{c_2^4}\ h\a^2 q\ \ha_x\,,
\end{split}
\eqlabel{fl3}
\end{equation}
\begin{equation}
\begin{split}
0=&H_{yy}''+\frac 12 H_{yy}'\ \left[\ln\frac{c_1^5c_2}{c_3^2}\right]'+
\frac 12\  H_{xx}'\ \left[\ln \frac {c_2}{c_1}\right]'
+\frac{c_3^2}{2c_1^2}\biggl(\w^2(H_{yy}-H_{xx})+q^2\frac{c_1^2}{c_2^2}( H_{tt}-H_{xx})\\
&+2\w q H_{ty}\biggr)
-\frac{c_3^2}{c_2^4}\ h^2\a^4\ (H_{xx}+H_{yy})+2\ \frac{c_3^2}{c_2^4}\ h\a^2 q\ \ha_x\,,
\end{split}
\eqlabel{fl4}
\end{equation}
\begin{equation}
\begin{split}
0=&\ha_x''+\ha_x'\ \left[\ln\frac {c_1}{c_3}\right]'+
\frac{c_3^2}{c_1^2}\ \ha_x\left(\w^2-\frac{c_1^2}{c_2^2} q^2\right)
+ \frac{c_3^2}{2c_2^2}\ h\a^2\ \biggl(q(H_{tt}+H_{xx}+H_{yy})\\
&+2 \w\ \frac{c_2^2}{c_1^2}\ H_{ty}\biggr)\,. 
\end{split}
\eqlabel{fl5}
\end{equation}
The number of the second order equations, of course, coincides 
with the number of the independent fluctuations.
Additionally, there are three first order constraints 
associated with the (partially) fixed diffeomorphism invariance 
\begin{equation}
\begin{split}
0=&\w\left(\left[H_{xx}+H_{yy}\right]'+
\left[\ln\frac{c_2}{c_1}\right]'\ (H_{xx}+H_{yy})\right)+
q\left(H_{ty}'+2\left[\ln\frac{c_2}{c_1}\right]'\ H_{ty}\right)\,,
\end{split}
\eqlabel{const1}
\end{equation}
\begin{equation}
\begin{split}
0=&q\left(\left[H_{tt}-H_{xx}\right]'-
\left[\ln\frac{c_2}{c_1}\right]'\ H_{tt}\right)+\frac{c_2^2}{c_1^2}\w\ H_{ty}'
+4\ h\a^2\ \frac{\ha_x'}{c_2^2}\,,
\end{split}
\eqlabel{const2}
\end{equation}
\begin{equation}
\begin{split}
0=&[\ln c_1c_2]'\left[H_{xx}+H_{yy}\right]'-[\ln{c_2^2}]'\ H_{tt}'+\frac{c_3^2}{c_1^2}
\biggl(\w^2\ (H_{xx}+H_{yy})+2\w q\ H_{ty}\\
&+q^2\ \frac{c_1^2}{c_2^2}\left(H_{tt}-H_{xx}\right)\biggr)
-2\frac{c_3^2}{c_2^4}\ h^2\a^4\ (H_{xx}+H_{yy})+4 \frac{c_3^2}{c_2^4}\ h\a^2q\ \ha_x\,.
\end{split}
\eqlabel{const3}
\end{equation}
The constraints are just the Einstein's equations obtained by varying the action 
with respect to the pure gauge metric components $h_{t r}, h_{y r}$ and $h_{r r}$. 
We explicitly verified that eqs.~\eqref{fl1}-\eqref{fl5} 
are consistent with the constraints~\eqref{const1}-\eqref{const3}. 

Now we introduce the fluctuations invariant under the residual
diffeomorphisms and gauge transformations preserving the gauge~\eqref{gaugec}.
Since we have five second-order equations and three constraints there must 
be two gauge invariant fluctuations. We find them to be
\begin{equation}
\begin{split}
Z_H=&4\frac{q}{\w} \ H_{ty}+2\ H_{yy}-
2 H_{xx}\left(1-\frac{q^2}{\w^2}\frac{c_1'c_1}{c_2'c_2}\right)+2\frac{q^2}{\w^2}
\frac{c_1^2}{c_2^2}\ H_{tt}\,,\\
Z_A=&\ha_x+\frac{1}{2q}\ h\a^2\ \left(H_{xx}-H_{yy}\right)\,.
\end{split}
\eqlabel{physical}
\end{equation}
Then 
from eqs.~\eqref{fl1}-\eqref{fl5} and~\eqref{const1}-\eqref{const3} 
we obtain two decoupled (gauge invariant) equations of motion for $Z_H$ and
$Z_A$\footnote{To achieve the decoupling 
one has to use the background equations of motion, 
\ie, the decoupling occurs only on-shell.}
\begin{equation}
\begin{split}
0=&A_HZ_H''+B_HZ_H'+C_HZ_H+D_HZ_A'+E_HZ_A\,, 
\end{split}
\eqlabel{zH}
\end{equation}
\begin{equation}
\begin{split}
0=&A_AZ_A''+B_AZ_A'+C_AZ_A+D_AZ_H'+E_A Z_H\,. 
\end{split}
\eqlabel{za}
\end{equation}
The connection coefficients $\{A_H,\cdots,E_{A}\}$ 
can we computed from~\eqref{fl1}-\eqref{fl5}, \eqref{const1}-\eqref{const3}
and \eqref{physical} using explicit expressions for the $c_i$'s, see~\eqref{metric}. 
Since these coefficients are very cumbersome
we will not present them in the paper.\footnote{The precise form 
of the equations~\eqref{zH} and~\eqref{za} is available from the authors upon request.}
In the next subsection, we will present 
the explicit form of the equations~\eqref{zH} and~\eqref{za} 
in the limit of small $\omega$ and $q$. This will be sufficient 
for our purposes.


\subsection{Boundary Conditions, Hydrodynamic Limit  and 
the Sound Wave Dispersion Relation}


\subsubsection{Boundary Conditions}


According to the general prescription~\cite{Andrei1, Andrei2}, in order
to obtain the dispersion relation (poles in the retarded Green's functions)
we have to impose the following boundary 
conditions on the 
gauge invariant fluctuations $\{Z_H, Z_A\}$.
\begin{itemize}
\item
$\{Z_H, Z_A\}$ must have
incoming wave boundary conditions 
near the horizon (as $r\to 1$);
\item $\{Z_H, Z_A\}$ must 
be normalizable near the boundary (as $r\to 0$).
\end{itemize}
The second condition is imposed because  coefficients of  non-normalizable solutions
appear as  poles in the retarded Green's functions.
Thus, setting them to zero will produce the dispersion relation. 
See~\cite{Andrei2} for details. 

Since the solution of interest is an incoming wave at the horizon it has the following 
general structure
\begin{equation}
Z_H(r)=\left[f(r)\right]^{\b_1}z_H(r)\,,\qquad Z_A(r)=\left[f(r)\right]^{\b_2}z_A(r)\,,
\eqlabel{redefz}
\end{equation}
where $f(r)$ is given by eq.~\eqref{metric} and the functions 
$z_H(r)$ and $z_A(r)$ are non-singular at the horizon. 
Then denoting
\begin{equation}
\lim_{r\to 1_-}z_H(r)\to z_{H}^{(0)}\ne 0\,,\qquad \lim_{r\to 1_-}z_A(r)\to z_{A}^{(0)}\ne 0\,,
\eqlabel{limboundary}
\end{equation}
we find from eq.~\eqref{zH} and eq.~\eqref{za} that as $x\equiv 1-r\to 0_+$
the following two equations must be satisfied
\begin{equation}
\begin{split}
&0=\a \ww^2 (h^2-3)^4 (h^2 \qq^2-3 \qq^2+4 \ww^2)^2 (4 \b_1^2+\ww^2)\ z_H^{(0)}\ \times \left(1+\calo(x)\right)\\
&+8 (h^2-3)^4 (h^2 \qq^2-3 \qq^2+4 \ww^2)^2 
(-h^2 \qq^2+3 \qq^2+16 \b_2-4 \ww^2) \qq h x \left((3-h^2) x\right)^{\b_2-\b_1} z_A^{(0)}\ 
\\
&\times \left(1+\calo(x)\right)\,,
\end{split}
\eqlabel{hor1}
\end{equation}
and
\begin{equation}
\begin{split}
&0=2 \a h \ww^2 (h^2-3)^3 (h^2 \qq^2-3 \qq^2+4 \ww^2) (-h^2 \qq^2+3 \qq^2+8 \b_1) x 
\ z_H^{(0)}\ \times \left(1+\calo(x)\right)\\
&+\qq (h^2-3)^4 (h^2 \qq^2-3 \qq^2+4 \ww^2)^2 (\ww^2+4 \b_2^2) \left((3-h^2) x\right)^{\b_2-\b_1}\ z_A^{(0)}\ 
\times \left(1+\calo(x)\right)\,,
\end{split}
\eqlabel{hor2}
\end{equation}
where we have defined $\ww = \omega/(2\pi T)$ and $\qq = q/(2\pi T)$.
From eqs.~\eqref{hor1} and~\eqref{hor2}, the
existence of a nontrivial solution to~\eqref{zH} and~\eqref{za} 
with incoming wave boundary 
conditions implies 
\begin{equation}
\begin{split}
0=&\alpha\ww^2 \qq ((3-h^2)x)^{\b_2-\beta_1}(3-h^2)^8(\qq^2 h^2+4\ww^2-3\qq^2)^4\\
&\times
\biggl\{(\ww^2+4\beta_2^2)(\ww^2+4\beta_1^2)+\calo(x)\biggr\}\,.
\end{split}
\eqlabel{caseb}
\end{equation}
Eq.~\eqref{caseb} suggests the following critical exponents:
\begin{equation}
\b_1=\b_2=-i\ \frac \ww2\,.
\eqlabel{casea}
\end{equation}
A detailed supergravity analysis of the solutions confirms that \eqref{casea} is indeed the correct choice.

Since \eqref{zH} and \eqref{za} are second order ODE's, each of them has two independent solutions.
Analyzing \eqref{zH}, \eqref{za} as $r\to 0$ shows that 
normalizability of $z_H$ implies that 
\begin{equation}
z_H(r)=\calo(r^3) \qquad {\rm as}\qquad r\to 0\,.
\eqlabel{zhb}
\end{equation}
While both solutions of $z_A$ are normalizable as $r\to 0$, requiring a fixed background magnetic field at the  boundary
implies that 
\begin{equation}
z_A(r)=\calo(r) \qquad {\rm as}\qquad r\to 0\,.
\eqlabel{zab}
\end{equation}


\subsubsection{Hydrodynamic Limit with $h\propto \qq$}


The aim of this subsection is to find the holographic solution 
dual to hydrodynamics with the magnetic field $B \sim q$. 
For this we have to 
study physical fluctuation equations \eqref{zH} and \eqref{za}, subject to the 
boundary conditions \eqref{casea}  and \eqref{zhb}, \eqref{zab} in the hydrodynamic approximation,
$\ww\to 0$, $\qq\to 0$ with $\frac{\ww}{\qq}$ and $\frac{h}{\qq}$ kept constant. 

To facilitate the hydrodynamic scaling we parametrize 
\begin{equation}
h\equiv |\qq| H\,.
\eqlabel{h}
\end{equation}
Furthermore, we parametrize
the sound quasinormal mode dispersion relation as follows
\begin{equation}
\ww=c_s\ \qq-\frac i2\biggl(\Gamma\ \qq^2+\Gamma_h\ h^2\biggr)+\calo\left(\qq^3,\qq h\right)\,,
\eqlabel{wpar}
\end{equation}
with $\{c_s,\Gamma,\Gamma_h\}$ kept fixed in the hydrodynamic scaling.
Without loss of generality we can assume that 
$\Gamma$ does not depend on $H^2$, 
while $\Gamma_h=\Gamma_h(H^2)$. We will look for a solution 
as a power expansion in $\qq$ and 
introduce
\begin{equation}
z_H=z_{H,0}+i\ \qq\ z_{H,1}+\calo(\qq^2)\,,\qquad z_A=h\left(z_{A,0}+i\ \qq\ z_{A,1}+\calo(\qq^2)\right)\,.
\eqlabel{hydroexp}
\end{equation}
Notice the $\calo(h)$ scaling of $z_A$ in the hydrodynamic limit. 
This is motivated by the field theory analysis. If $\rho=0$,
the $x$ components of 
the current in~\eqref{2.20} is proportional to the magnetic field.
Since the equations for $z_{H}$ and $z_A$ are homogeneous
we can rescale $z_H(r)$ and $z_A(r)$ to make $z_{H}(r)$
be equal to unity on the horizon. 
It is convenient to make a choice that the leading solution $z_{H, 0}$ equals to unity 
and all the subleading contributions $z_{H, 1}(r), \dots$
vanish on the horizon. 
That is, we impose that
\begin{equation}
\lim_{r\to 1_-}z_H(r)=1\,,\qquad 
\lim_{r\to 1_-}z_{H,0}=1\,,\quad  \lim_{r\to 1_-}z_{H,1}=0\,.
\eqlabel{bc1}
\end{equation}

To leading order in the hydrodynamic approximation, and subject to the appropriate boundary conditions,
we find from \eqref{zH} and \eqref{za} that
\begin{equation}
\begin{split}
0=&z_{H,0}''-
\biggl(\left(9 (4 c_s^2+r^3-4)^2+64 r^2 (r-1) (r^2+r+1) (4 c_s^2-3) H^2\right)\\
&\times (r^2+r+1) (r-1) r\biggr)^{-1}\ \biggl(9 (4-4 c_s^2-r^3) (4 r^3 c_s^2+8 c_s^2-8+4 r^3-5 r^6)\\
&+128 r^2 (r-1)^2 (r^2+r+1)^2 (4 c_s^2-3) H^2\biggr)\ z_{H,0}'
+\biggl((9 (4 c_s^2+r^3-4)^2\\
&+64 r^2 (r-1) (r^2+r+1) (4 c_s^2-3) H^2) (r^2+r+1) (r-1)\biggr)^{-1}\\
&\times 81 (4 c_s^2+r^3-4) r^4\  z_{H,0}\,,
\end{split}
\eqlabel{res11} 
\end{equation}
and
\begin{equation}
\begin{split}
0=&z_{H,0}'+\frac{3 r^2 }{4 c_s^2-2-r^3}\ z_{H,0}\,.
\end{split}
\eqlabel{res12} 
\end{equation}
Solving \eqref{res12} subject to the boundary conditions \eqref{bc1} implies that
\begin{equation}
z_{H,0}=r^3\,,\qquad c_s=\pm \frac{1}{\sqrt{2}}\,.
\eqlabel{sol1res1}
\end{equation}
Given \eqref{sol1res1}, it is straightforward to verify that  \eqref{res11} is satisfied as well. 
Thus, we have solved the equations of motion to leading order in the hydrodynamic approximation. 

Now we move to next-to-leading order.
Using \eqref{sol1res1}, we find from \eqref{zH} and \eqref{za}
the following two equations
\begin{equation}
\begin{split}
0=&z_{H,1}''-\frac{128 r^2 (r-1)^2 (r^2+r+1)^2 H^2-9 (r^3-2) (5 r^6-6 r^3+4)}{r \Delta_3 (r^2+r+1) (r-1)}\ z_{H,1}'\\
&+\frac{576 i\ (r^3-2) H^2 r^4}{\alpha \Delta_3}\ z_{A,0}'-
\frac{81 (r^3-2) r^4}{\Delta_3 (r^2+r+1) (r-1)}
\  z_{H,1}\\
&+\frac{9 \sqrt 2 (2 H^2 (9 \Gamma_h-8 r^2)-9+18 \Gamma) (r^3-2) r^4}{\Delta_3 (r^2+r+1) (r-1)}\,,
\end{split}
\eqlabel{res13}
\end{equation}
and
\begin{equation}
\begin{split}
0=&z_{A,0}''+
\frac{r (64 (r^3-1) (r^3+2) H^2-27 r (r^3-2)^2)}{\Delta_3(r^3-1)}\
z_{A,0}'
+\frac{6 i\ r^2 \a}{\Delta_3}\ z_{H,1}'-\frac{18 i\ r \a}{\Delta_3 }\ z_{H,1} \\
&+\frac{9 i\ r \sqrt2 \a (8 \Gamma_h H^2 
(r^3-1)+8 r^3 \Gamma-r^6-8 \Gamma)}{2 \Delta_3  (r^3-1)}\,,
\end{split}
\eqlabel{res14}
\end{equation}
where 
\begin{equation}
\Delta_3=64 r^2 (r-1)(r^2+r+1)H^2-9(r^3-2)^2\,.
\eqlabel{defdelta3}
\end{equation}
It is difficult to solve eqs.~\eqref{res13} and \eqref{res14} analytically for all values of $H^2$. 
However, for our purposes it is enough to find a solution to order $H^2$.
Expanding $z_{H, 1}$, $z_{A,1}$ and $\Gamma_h$
perturbatively in $H^2$,
\begin{equation}
\begin{split}
&z_{H,1}=z_{H,1,0}+H^2 z_{H,1,1}+\calo(H^4)\,,\qquad z_{A,0}=z_{A,0,0}+H^2 z_{A,0,1}+\calo(H^4)\,,\\
&\Gamma_h=\Gamma_{h,0}+\calo(H^2)\,,
\end{split}
\eqlabel{hexpand}
\end{equation}
we find that the solution 
satisfying the boundary conditions stated in eqs.~\eqref{zhb}, \eqref{zab} 
and~\eqref{bc1} is given by
\begin{equation}
\begin{split}
z_{H,1,0}=&0\,,\qquad z_{H,1,1}=0\,,\qquad z_{A,0,1}=0\,,\\
z_{A,0,0}=&-\frac{i\ \sqrt 2\alpha(9\ln(r^2+r+1)+6\sqrt3\arctan\left(\frac{2r+1}{\sqrt{3}}\right)-\pi \sqrt3)}
{72}\,,\\
&\Gamma=\frac 12\,,\qquad \Gamma_{h,0}=\frac 89\,.
\end{split}
\eqlabel{solveHexp}
\end{equation}
Here the last line comes from requiring that $z_{H, 1}$ vanishes on the boundary. 


\subsubsection{Comparison with Field Theory}


To summarize the results, we have obtained the following dispersion relation 
\begin{equation}
\ww=\pm \frac{1}{\sqrt{2}} \ \qq-\frac i2
\biggl(\frac{1}{2} \qq^2+\frac{8}{9} \ H^2\qq^2\biggr)
+\calo\left(\qq^3,\qq^2 H\right)\,.
\eqlabel{m1}
\end{equation}
Let us compare it with the field theory counterpart~\eqref{2.38}. 
For this we will rewrite eq.~\eqref{m1} in notation of~\eqref{2.38}. 
First, we will recall that 
\begin{equation}
\ww=\frac{\omega}{2 \pi T}\,, \quad \qq=\frac{q}{2 \pi T}\,. 
\eqlabel{m2}
\end{equation}
Second, we will express the variables $(H, \alpha)$ which are natural to use on the 
supergravity side in terms of $(b, T)$. We get 
\begin{equation}
H=\frac{2 \pi T b}{\alpha^2} =\frac{2 \pi T b}{\alpha_0^2}+{\cal O}(b^3 q^2)\,, 
\eqlabel{m3}
\end{equation}
where 
\begin{equation}
\alpha_0=\frac{4 \pi T}{3}\,. 
\eqlabel{m4}
\end{equation}
Note that the relation is non-linear because $\alpha$ also depends on $B=b q$. 
However, for our purposes it is sufficient to take it to leading order. 
Substituting eqs.~\eqref{m2}-\eqref{m4} into eq.~\eqref{m1} we obtain
(up to higher order terms)
\begin{equation}
\ww=\frac{1}{\sqrt{2}} \ \qq-\frac i2
\biggl(\frac{\qq^2}{4 \pi T} +\frac{4}{3}\ \frac{b^2 \qq^2}{\alpha_0^3}\biggr)\,.
\eqlabel{m5}
\end{equation}
Comparing it with eq.~\eqref{2.38}, first, we find that 
\begin{equation}
\frac{\eta}{\epsilon+P} =\frac{1}{4 \pi T}\,. 
\label{m6}
\end{equation}
This result agrees with earlier calculations 
of the shear viscosity $\eta$ in~\cite{Herzog1, Herzog2}. 
Second, we obtain
\begin{equation}
\frac{\sigma_Q}{\epsilon+P}=\frac{4}{3 \alpha_0^3}\,.
\label{m7}
\end{equation}
Recalling results for $\epsilon$ and $P$ from subsection 3.1 we conclude that 
\begin{equation}
\sigma_Q=\frac{1}{g^2}\,. 
\label{m8}
\end{equation}
Thus, we reproduced the result for the conductivity coefficient (to leading order 
in the magnetic field) obtained earlier in~\cite{Kovtun, HH}. 

We see that we have a perfect agreement with our field theory analysis
and with the earlier calculations of the transport coefficients. 


\subsubsection{Hydrodynamic Limit with $h\propto \sqrt{|\qq|}$}


In this subsection, we will study the holographic solution dual
to hydrodynamics with the magnetic field $B \sim \sqrt{q}$. 
For this we will 
consider the physical fluctuation equations~\eqref{zH} and~\eqref{za}, subject to the 
boundary conditions~\eqref{casea}   and~\eqref{zhb}, \eqref{zab} 
in the hydrodynamic approximation,
$\ww\to 0$, $\qq\to 0$ with $\frac{\ww}{\qq}$ and $\frac{h}{\sqrt{|\qq|}}$ kept constant. 

The analysis will be similar to the one in subsection 3.3.2.
To facilitate the hydrodynamic scaling we parametrize 
\begin{equation}
h\equiv \sqrt{|\qq|} H\,.
\eqlabel{hs}
\end{equation}
The sound quasinormal mode dispersion relation is parametrized as follows
\begin{equation}
\begin{split}
\ww=&\Delta\ h^2+c_s\ \qq-\frac i2\biggl(\Gamma\ \qq^2+
\Gamma_h\ h^4\biggr)+\calo\left(\qq^3,\qq^2 h^2,\qq h^4, h^6\right)\\
=&\Gamma_0\ \qq+i \Gamma_1\ \qq^2+\calo(\qq^3)\,,\\
\Gamma_0\equiv& \Delta H^2+c_s\,,\qquad \Gamma_1=-\frac 12 \biggl(\Gamma+\Gamma_h H^4\biggr)\,,  
\end{split}
\eqlabel{wpars}
\end{equation}
with $\{\Delta,c_s,\Gamma,\Gamma_h\}$ kept fixed in the hydrodynamic scaling.
We look for a solution 
as a series in $\qq$,
\begin{equation}
z_H=z_{H,0}+i\ \qq\ z_{H,1}+\qq^2\ z_{H,2}+\calo(\qq^3)\,,
\qquad z_A=h\left(z_{A,0}+i\ \qq\ z_{A,1}+\qq^2 z_{A,2}+\calo(\qq^3)\right)\,,
\eqlabel{hydroexps}
\end{equation}
and impose (without loss of generality) that 
\begin{equation}
\lim_{r\to 1_-}z_H(r)=1\qquad \Rightarrow\qquad  
\lim_{r\to 1_-}z_{H,0}=1 \ \&\ \lim_{r\to 1_-}z_{H,1}=\lim_{r\to 1_-}z_{H,2}=0\,.
\eqlabel{bc1s}
\end{equation}

To leading order in the hydrodynamic approximation we obtain from \eqref{zH}: 
\begin{equation}
\begin{split}
0=&z_{H,0}''-\frac 2r\ z_{H,0}' \,,
\end{split}
\eqlabel{res11s} 
\end{equation}
which gives rise to the solution 
\begin{equation}
z_{H,0}=r^3\;.
\eqlabel{fff}
\end{equation}
Using this solution for $z_{H,0}$, we find from \eqref{za} the following equation 
for $z_{A, 0}$
\begin{equation}
\begin{split}
0=&z_{A,0}''+\frac{r^3+2}{(r^3-1) r}\ z_{A,0}'
+\frac{9\Gamma_0^2 (2\Gamma_0^2-1)\alpha}{8(r^3-1)(4\Gamma_0^2-3)H^2r}\,.
\end{split}
\eqlabel{res12s} 
\end{equation}
Solving \eqref{res12s} subject to the boundary conditions~\eqref{zab} gives
\begin{equation}
\begin{split}
z_{A,0}=&\frac{(2\Gamma_0^2-1) \Gamma_0^2\a}{32 H^2 (4 \Gamma_0^2
-3)}
\biggl(\pi \sqrt3 -6 \sqrt3 \arctan\frac{1+2 r}{\sqrt 3}-9 \ln(r^2+r+1)\biggr)\;.
\end{split}
\eqlabel{sol1res1s}
\end{equation}

This finishes our consideration of the equations of motion to leading order. 
Note that we were not able to determine any of the transport coefficients 
in the dispersion relation~\eqref{wpars}. Moving on to 
next-to-leading order in the hydrodynamic approximation we find from
\eqref{zH} and \eqref{za} 
\begin{equation}
\begin{split}
0=&z_{H,1}''-\frac2r\ z_{H,1}'+\frac{9i\ r^2 (r^3+4 \Gamma_0^2-4)}
{2\Gamma_0^2 \a(r^3-1)}\ z_{A,0}' 
- \frac{i\ r^2}{32 (r^3-1)^2 (4 \Gamma_0^2-3) H^2} \biggl(324-81 r^3\\
&+432 i\ (r^3-2) H^2 r^2 
\Gamma_0+(162 r^3+512 H^4 r^6-972-512 H^4 r^3) \Gamma_0^2\\
&-576 i\ (r^3-2) H^2 r^2 \Gamma_0^3+648 \Gamma_0^4\biggr)\,,
\end{split}
\eqlabel{res13s}
\end{equation}
and
\begin{equation}
\begin{split}
0=&z_{A,1}''+\frac{r^3+2}{r (r^3-1)}\ z_{A,1}'
+\frac{3\Gamma_0^2 (4 \Gamma_0^2-2-r^3) \a}{16H^2 r^3 (r^3-1) 
(4 \Gamma_0^2-3)} z_{H,1}'\\
&+ \frac{9\Gamma_0^2 \a}
{16r H^2 (r^3-1) (4 \Gamma_0^2-3)} z_{H,1}
+J_{source}\biggl[z_{A,0}(r);\{\Gamma_0,\Gamma_1,H\};r\biggr]\,,
\end{split}
\eqlabel{res14s}
\end{equation}
where the source term $J_{source}$ is a linear functional of 
$z_{A,0}$ also depending on the transport coefficients
and the magnetic field , 
$\{\Gamma_0,\Gamma_1,H\}$, as well as explicitly on  
$r$.\footnote{The explicit 
expression for $J_{source}$ is extremely lengthy and cumbersome and we find 
it meaningless to put it in the paper. 
It is available from the authors upon request.}

It is straightforward to analyze the asymptotic 
solution to~\eqref{res13s} near the horizon, \ie, as $x\equiv 1-r\to 0_+$. 
For generic values of $\Gamma_0$ we find 
that $z_{H,1}$ has a simple pole and a logarithmic singularity near the horizon. Namely,
\begin{equation}
\begin{split}
z_{H,1}=&\calc\ \biggl\{
-\frac{i}{32 H^2}\  \ln x\biggr\}
+{\rm finite}\,,
\end{split}
\eqlabel{horzh1}
\end{equation}
where
\begin{equation}
\calc=18 \Gamma_0^2+16i\ H^2\Gamma_0-9\,.
\eqlabel{defc}
\end{equation}
Regularity at the horizon implies that $\calc=0$. It leads to the following solution
\begin{equation}
\Gamma_0=c_s+\Delta H^2=-i\ \frac 49 H^2\pm \frac {1}{\sqrt 2}\sqrt{1-\frac{32}{81}H^4}\,. 
\eqlabel{dcs}
\end{equation} 
Without loss of generality, we can assume that $c_s$ is independent of 
$H^2$, while all the $H^2$ dependence resides in $\Delta$. Hence, we obtain
\begin{equation}
c_s=\pm \frac{1}{\sqrt 2}\,,\qquad \Delta_{\pm}=-i\ \frac 49\mp \frac{1}{H^2\sqrt{2}}\ 
\left(1-\sqrt{1-\frac{32}{81}H^4}\right)\,,
\eqlabel{resdcs}
\end{equation} 
where the signs $\pm$ are correlated in the above expressions.
Given \eqref{resdcs}, a nonsingular function $z_{H,1}$ subject to the 
boundary conditions \eqref{zhb}, \eqref{bc1s}
is uniquely specified:
\begin{equation}
\begin{split}
z_{H,1}=&\frac{2 i\ H^2 (1-r)(16H^2\Gamma_0+9 i)}{27 (32H^2\Gamma_0-9 i)}
\ \times\ \biggl((1+r+r^2) (12 \sqrt 3 \arctan\frac{1+2 r}{\sqrt 3}\\
&+18 \ln(1+r+r^2))
-2 \sqrt 3 \pi-(36+2\sqrt{3}\pi)( r+r^2)-27 r^3
\biggr)\,.
\end{split}
\eqlabel{zh1res}
\end{equation}

Now we move on to eq~\eqref{res14s}.
Its general solution is of the form
\begin{equation}
\begin{split}
z_{A,1}'=&\frac{r^2}{r^3-1}\biggl(\int_0^rd\xi\ \frac{(1-\xi^3)\ \hat{J}_{source}(\xi)}{\xi^2}+\calc_{A,1,1}\biggr)
\end{split}
\eqlabel{za1s}
\end{equation}
where $\calc_{A,1,1}$ is an arbitrary integration constant and 
\begin{equation}
\begin{split}
\hat{J}_{source}(r)\equiv &J_{source}\biggl[z_{A,0}(r);\{\Gamma_0,\Gamma_1,H\};r\biggr]
+\frac{3\Gamma_0^2 (4 \Gamma_0^2-2-r^3) \a}{16H^2 r^3 (r^3-1) 
(4 \Gamma_0^2-3)} z_{H,1}'
\\
&+ \frac{9\Gamma_0^2 \a}
{16r H^2 (r^3-1) (4 \Gamma_0^2-3)} z_{H,1}
\end{split}
\eqlabel{jhat}
\end{equation}
Since 
\begin{equation}
\begin{split}
&\hat{J}_{source}\propto \frac 1r\,,\qquad {\rm as}\qquad r\to 0_+\,,\\
&\hat{J}_{source}\propto \frac {1}{x}\,,\qquad {\rm as}\qquad x=1-r\to 0_+\,,
\end{split}
\eqlabel{assza1s1}
\end{equation}
it is clear from \eqref{za1s} that for any value
of $\Gamma_1$ we can adjust the integration constant $\calc_{A,1,1}$ to remove singularity of 
$z_{A,1}$ at the horizon, $x\to 0_+$; the second constant (obtained from integrating \eqref{za1s}) 
can be fixed to insure that $z_{A,1}$ vanishes as $r\to 0$, see \eqref{zab}.   
Thus, we cannot determine $\Gamma_1$ at this order. This is not surprising, 
given that to determine the leading order transport coefficient $\Gamma_0$ 
we had to consider the $\calo(\qq)$ order in the hydrodynamic approximation of $z_H$.

In order to determine $\Gamma_1$ we have to consider $\calo(\qq^2)$ in the hydrodynamic approximation for 
$z_{H}$. The equation of motion 
for $z_{H,2}$ takes the following form
\begin{equation}
\begin{split}
0=&z_{H,2}'''-\frac2r\ z_{H,2}'-\frac{9i\ r^2 (r^3+4 \Gamma_0^2-4)}
{2\Gamma_0^2 \a(r^3-1)}\ z_{A,1}' 
+\cali_{source}\biggl[z_{A,0},z_{H,1};\{\Gamma_0,\Gamma_1,H\};r\biggr]\,,
\end{split}
\eqlabel{zh2}
\end{equation}
where $\cali_{source}$ is a new source term.\footnote{Its explicit expression 
is too lengthy and cumbersome to be put in the paper. It is available from the
authors upon request.}
Although we can not solve analytically for $z_{H,2}$, it is straightforward to 
construct a power series solution, first, for $z_{A,1}$ and then for $z_{H,2}$
near the horizon, $x=1-r\to 0_+$. Much like in~\eqref{horzh1}, we find that 
$z_{H,2}$ is non-singular at the horizon, provided
\begin{equation}
\begin{split}
\Gamma_1=&-\frac 14-\frac 89H^4\pm\frac{2iH^2(32 H^4-45)}{9\sqrt{162-64 H^4}}
\end{split}
\eqlabel{resdd1}
\end{equation} 
where $\pm$ sign correlates with the corresponding sign in \eqref{dcs}.
Perturbatively in $H^2$,
\begin{equation}
\begin{split}
\Gamma_1=&-\frac14\mp\frac{i\ 5\sqrt 2}{9}H^2+\calo(H^4)\,.
\end{split}
\eqlabel{ddpert}
\end{equation}


\subsubsection{Comparison with Field Theory}


To summarize computations performed above, we have obtained the following sound wave
\begin{equation}
\begin{split}
\ww & = \qq \left(-\frac{4 i}{9}H^2 \pm \frac{1}{\sqrt{2}}\sqrt{1-\frac{32}{81}H^4}\right) 
+ i \qq^2 \left(-\frac{1}{4}\mp\frac{i\ 5\sqrt 2}{9}H^2+ 
{\cal O}(H^4)\right)\,. 
\end{split}
\eqlabel{mm1}
\end{equation}
We would like to compare it with field theory. First, we will consider terms
of order $q$ and compare them with the field theory counterpart~\eqref{2.41}.
Following the same logic as in subsection 3.3.3 it is straightforward to check 
that terms of order $q$ are in total agreement with eq.~\eqref{2.41} if, as before, 
\begin{equation}
\frac{\eta}{\epsilon+P}=\frac{1}{4 \pi T}\,, \quad \sigma_Q=\frac{1}{g^2}\,. 
\eqlabel{mm2}
\end{equation}
Thus, we have obtained an agreement with field theory at leading order in $\qq$. 
Note that it holds to all orders in $H$.  

Let us now compare terms of order $q^2$ in
eq.~\eqref{mm1} with the corresponding field theory prediction~\eqref{2.44}. 
It is easy to check (it has already been done in subsection 3.3.3) 
that the results agree for vanishing magnetic field. 
However, we have a disagreement at order $q^2 H^2$. 
The field theory result at this order is 
\begin{equation}
\ww \sim \mp\sqrt{2}\frac{\qq^2 H^2}{9}\,, 
\eqlabel{mm3}
\end{equation}
where eqs.~\eqref{mm2} have been taken into account. 
On the contrary, the supergravity result is 
\begin{equation}
\ww \sim \pm \frac{\ 5\sqrt 2}{9}\qq^2 H^2\,. 
\eqlabel{mm4}
\end{equation}
We have been unable to identify the source of this disagreement in the framework of HKMS viscous 
magneto-hydrodynamics. It is natural to expect that introduction of additional transport coefficients 
would resolve this puzzle.  We hope to discuss this issue in more details 
in \cite{inprep}.


\section{Conclusion}


In this paper we discussed propagation of the sound waves in magnetic fluids in (2+1) dimensions. 
We used the strongly coupled $M2$-brane plasma and the general 
setting of the holographic gauge theory/string 
theory duality to test relativistic viscous magneto-hydrodynamics 
of HKMS \cite{Kovtun}. We found that HKMS magneto-hydrodynamics,
generalizing the arguments of Landau and Lifshitz~\cite{LL} 
in constructing dissipative entropy currents,
adequately describes propagation of sound modes to the order in the hydrodynamic 
limit first sensitive to the external 
field. There is, however, a disagreement in next order in the hydrodynamic 
approximation (still in the context of the 
first-order hydrodynamics). Such a disagreement suggests that 
 additional 
transport coefficients, beyond those introduced in~\cite{Kovtun}, might be needed to 
describe dissipation in (2+1) 
magnetic fluids even to first order in the local velocity gradients.  

Finally, it would be interesting to analyze shear modes in (2+1) dimensional strongly 
coupled magnetic plasma. While the general
theorem \cite{genshear} guarantees that the shear viscosity 
attains its universal value \cite{u1,u2,u3}, a deeper understanding 
of the propagation of the shear modes might help in constructing a 
general theory of relativistic viscous magneto-hydrodynamics.
   

\section*{Acknowledgments}

We would like to thank Colin Denniston, Sean Hartnoll, Chris Herzog,  Pavel Kovtun and Rob Myers
for valuable discussions.
Research at Perimeter Institute is supported by the
Government of Canada through Industry Canada and by the Province of
Ontario through the Ministry of Research \& Innovation. AB
gratefully acknowledges further support by an NSERC Discovery grant
and support through the Early Researcher Award program by the
Province of Ontario. 


\appendix
\section{Bulk Diffeomorphisms and Boundary Weyl Transformations}


In this section, we  
show that the holographic stress-energy tensor is indeed 
traceless\footnote{Up to a possible conformal anomaly, which is known to be zero in odd dimensions.}. 
For the sake of generality, we will work in a bulk of an arbitrary dimension
$d+1$. 
We write the metric using the Fefferman-Graham coordinates \cite{FG} as follows
\be 
\label{metricFG} 
ds^2 = \frac{1}{r^2} \left( dr^2+ g_{\mu \nu}(r,x) dx^\mu dx^\nu  \right)\;.
\ee
The boundary condition on the metric is
\be 
\lim_{r \rightarrow 0}  g_{\mu \nu}(r,x)= g_{\mu \nu}^{(0)}\;,
\ee
where $g_{\mu \nu}^{(0)}$ is the 
background metric which the gauge theory is defined.
We do not need to worry about the exact nature of the sub-leading terms.

The bulk diffeomorphisms that preserve the form of the metric (\ref{metricFG}) satisfy the Killing's equations
\be 
\label{killing}
\nabla_r \xi_r = 0\;,\;\;\; \nabla_{(\mu} \xi_{r)} = 0\;.\ee
We are interested in bulk diffeomorphisms that generate a 
Weyl transformation in the boundary metric $g^{(0)}_{\mu \nu}$. 
One can show that such solutions have the asymptotic behavior
\be\label{diff} \xi_r = \frac{ \sigma(x)}{r} + \ldots  \;,\;\;\; 
\xi_\mu = - \frac{1}{2} \partial_\mu \sigma(x) + \ldots \ee
It is straightforward to show that 
the transformation of the boundary metric 
under the diffeomorphism~\eqref{diff}
is given by
\bea
\label{xformg}\delta g_{\mu \nu}^{(0)} =\lim_{r\rightarrow 0 } 
\left( - {\cal L}_\xi g_{\mu \nu} \right)=   2 \sigma(x) g_{\mu \nu}^{(0)} \;.\eea

In order to compute the transformation of the boundary gauge field,  
we need to know the asymptotic behavior of the classical solutions of the bulk Maxwell's equations
\be \nabla_M F^{MN} = 0\;.\ee
In the gauge $A_r= 0$, Maxwell's equations take the form
\bea \label{meq1}&&\ddot A_\mu   + (2 -d) \dot A_\mu  + \frac{1}{2} g^{\alpha \beta} \dot g_{\alpha\beta} \dot A_\mu + g_{\mu \alpha} \dot g^{\alpha \beta} \dot A_\beta + \frac{r^2}{\sqrt{-g} }g_{\mu \alpha} \partial_\beta\left( \sqrt{-g} g^{\beta \gamma} g^{\alpha \delta} F_{\gamma \delta} \right) = 0\;, 
\nonumber\\ 
\\
 \label{meq2}&&\partial_\mu \left(\sqrt{-g} g^{\mu \nu} \dot A_\nu\right) = 0\;, \eea
where $\dot f = r\partial_r f$. One can then look for asymptotic solutions 
to these equations in the form $A_\mu \sim r^k A_\mu^{(0)}(x) + \ldots$. 
One finds that the exponent $k$ must satisfy
\be k(2-d  + k) = 0\;.\ee
Therefore, the leading term near the boundary is finite ($k = 0$)
\be \lim_{r\rightarrow 0} A_\mu(r,x) = A_\mu^{(0)}(x)\;,\ee
and we associate it with the boundary gauge field.

For the sake of completeness, let us mention that in the case of a massive bulk gauge 
field, the RHS of the first Maxwell equation (\ref{meq1}) equals $m^2 A_\mu$. 
Therefore the asymptotic behavior of the field is modified to
\be \lim_{r\rightarrow 0} A_\mu(r,x) =  r^k A_\mu^{(0)}(x)\;,\ee
where
\be \label{k} k = \frac{1}{2} \left[d-2 -\sqrt{(d-2)^2+4 m^2}\right]\;.\ee
We still identify $A_\mu^{(0)}(x)$ as the background gauge field in the field theory. 

We are now ready to study the transformation properties of the bulk gauge field under the diffeomorphisms (\ref{diff}). 
 Our gauge is $A_r = 0$. However, one needs to check that the 
diffeomorphisms (\ref{diff}) respect such gauge choice. 
That is, in general one might need to make a compensating gauge 
transformation to ensure $\delta A_r = 0$. The leading change in the $A_r$ component is
\be \lim_{r\rightarrow 0} \delta_\xi A_r = 
\lim_{r\rightarrow 0}(-{\cal L}_\xi A_r )  = r^{k + 1} A_\mu^{(0)} \partial^\mu \sigma(x)\;.\ee
We see that the compensating gauge transformation must be 
subleading at the boundary. More precisely, we take $\delta_\lambda A_M = \partial_M \lambda$, 
and from the requirement $(\delta_\xi  + \delta_\lambda)A_r = 0$ we find that
\be \lim_{r\rightarrow 0} \lambda =  - \frac{1}{k+2} r^{k+2} A_\mu^{(0)} \partial^\mu \sigma(x)\;.\ee
Finally, we find that the total asymptotic transformation of the gauge field in the $x^\mu$ directions is
\be \lim_{r\rightarrow 0} (\delta_\xi + \delta_\lambda) A_\mu = - k \sigma(x) A_\mu^{(0)}(x) r^{k} \;,\ee 
which translates to
\be \label{xformA} \delta A_\mu^{(0)} = - k \sigma(x) A_\mu^{(0)}(x)\;.\ee

Inserting the transformation rules (\ref{xformg}) and (\ref{xformA}) into the boundary action (\ref{2.1}), we obtain the Ward identity
\be \bra T^\mu_\mu\ket = k \bra J^\mu \ket A_\mu^{(0)}\;,\ee
where $k$ is given in eq. (\ref{k}) for a massive gauge field. 
For a massless gauge field ($k = 0$), which is the case of interest in this paper, we obtain a 
traceless stress-energy tensor.



\end{document}